\documentclass[journal]{IEEEtran}
\usepackage{graphicx} 
\usepackage[cmex10]{amsmath}
\DeclareMathOperator*{\argmax}{arg\,max}
\usepackage{algorithm}
\usepackage{algpseudocode}
\usepackage[version=4]{mhchem}
\usepackage{multirow}
\usepackage{ amssymb }

\usepackage{hyperref}
\usepackage{fancyref}
\usepackage[noadjust]{cite}

\title{Real-time, Adaptive Radiological Anomaly Detection and Isotope Identification Using Non-negative Matrix Factorization}
\author{A.\,C.~Jones,
        M.\,S.~Bandstra,
        S.~Faaland,
        Y.~Lai,
        N.~Abgrall,
        S.~Suchyta,
        and R.\,J.~Cooper
\thanks{A.C. Jones, M.S. Bandstra, S. Faaland, Y. Lai, N. Abgrall, and R.J.~Cooper, are with the Nuclear Science Division at Lawrence Berkeley National Laboratory, Berkeley, CA 94720 USA, e-mail: acjones@lbl.gov. S. Suchyta is with Nevada National Security Sites, Remote Sensing Laboratory - Andrews, Joint Base Andrews, MD 20762.}
\thanks{This work was performed under the auspices of the U.S. Department of Energy by Lawrence Berkeley National Laboratory (LBNL) under Contract DE-AC02-05CH11231. This work was done by Mission Support and Test Services, LLC, under Contract No. DE-NA0003624 with the U.S. Department of Energy and the National Nuclear Security Administration’s Office of Defense Nuclear Nonproliferation DOE/NV/03624\texttt{-{}-}2197.}}
\date{September 2025}
\IEEEpeerreviewmaketitle

\begin{document}

\maketitle
\begin{abstract}
Spectroscopic anomaly detection and isotope identification algorithms are integral components in nuclear nonproliferation applications such as search operations. The task is especially challenging in the case of mobile detector systems due to the fact that the observed gamma-ray background changes more than for a static detector system, and a pretrained background model can easily find itself out of domain. The result is that algorithms may exceed their intended false alarm rate, or sacrifice detection sensitivity in order to maintain the desired false alarm rate. Non-negative matrix factorization (NMF) has been shown to be a powerful tool for spectral anomaly detection and identification, but, like many similar algorithms that rely on data-driven background models, in its conventional implementation it is unable to update in real time to account for environmental changes that affect the background spectroscopic signature. We have developed a novel NMF-based algorithm that periodically updates its background model to accommodate changing environmental conditions. The Adaptive NMF algorithm involves fewer assumptions about its environment, making it more generalizable than existing NMF-based methods while maintaining or exceeding detection performance on simulated and real-world datasets.
\end{abstract}

\section{Introduction}
    Spectroscopic anomaly detection and isotope identification algorithms are integral components in nuclear nonproliferation applications such as detection of radioactive materials that are lost, stolen, or otherwise outside of regulatory control. Such algorithms have been used in applications including homeland security~\cite{kouzes, runkle, geel, preventing}; contamination mapping, monitoring, and remediation~\cite{ENGELBRECHT20201, fukushima}; and emergency response (e.g., recovery of lost sources)~\cite{australia, aerial4}. 
    
    Numerous methods have been developed to detect and identify anomalous sources of gamma radiation in different systems, including for networked arrays of detectors~\cite{chandy_networked_2008, cooper_intelligent_2012, panda2, nemzek_distributed_2004, hoteling_analysis_2021, networks, deb_iterative_2013, rao_network_2015, panda1, gnn}, radiation portal monitors~\cite{runkle_analysis_2006, boardman_principal_2012, ghelman_rpms_2013, rpm, cosofret_utilization_2013}, aerial detector systems~\cite{cresswell_137cs_2006,kock_background_2014,detwiler_spectral_2015, aerial1}, and ground-based mobile detector systems~\cite{mitchell_mobile_2009,penny_dual-sided_2011, zelakiewicz_sorisstandoff_2011, radmap1, tandon_detection_2016,miller_gamma-ray_2020}. Each of these data collection methods presents unique challenges in algorithm development. The main challenges of search tasks performed using mobile detector systems (the use case on which this paper will focus), particularly in complex environments such as cities, are the following:
    \begin{enumerate}
        \item \label{rapid} \textbf{Rapidly changing background gamma-ray radiation signatures}: Variations of concentrations of naturally occurring radioactive material (NORM) in building materials, density of buildings in a given area, visibility of the sky, and static and dynamic occlusion from objects in the environment are just a few factors that influence the temporally and spatially varying background spectrum. The ability to accurately account for changes in the background is key to achieving high detection sensitivity and controlling false positive rates~\cite{aucott_effects_2014,kock_background_2014,pfund_improvements_2016,stewart_understanding_2018}.
        \item \label{weak} \textbf{Weak/low signal-to-noise ratio (SNR) sources}: Effective algorithms often need to be able to detect weak anomalies, as shielding and other environmental attenuation factors will reduce the observable activity. The ability to detect weak anomalies is closely tied to maintaining an accurate background model. 
        \item \label{far} \textbf{Maintaining an operationally relevant false-alarm rate (FAR)}: From an operational standpoint, it is essential that detection and identification algorithms maintain a FAR at or below some predetermined level (e.g., 1~per~8~hours). False alarms waste time and resources as an operator will have to investigate the alarm. In order to maintain a low FAR, it is helpful for the alarm metric distribution to be stable and well-understood to make threshold setting simple. 
    \end{enumerate}

    To date, many approaches which use varying degrees of spectral information have been developed. Some algorithms use only the gross count rate of gamma rays registered in the detector~\cite{jarman_comparison_2008,archer_systematic_2015}, disregarding their energies. Others rely on truncated spectral regions of interest~\cite{mcew} or ratios between coarse spectral bins~\cite{detwiler_spectral_2015, pfund_improvements_2016}, making them better suited to detect weak anomalies (as in point~\ref{weak}, above) than algorithms using gross count rate alone. Algorithms that perform full-spectrum analysis~\cite{jaewon, lalor2024enhancingradioisotopeidentificationgamma, arad, boardman_principal_2012, sullivan_wavelet_2005, cosofret_utilization_2013} make use of all available spectroscopic information, which generally results in better detection performance. In recent years, various classical~\cite{bayes, jaewon, rpm} and deep learning~\cite{gnn, arad, nn2, kamuda_comparison_2020, lalor2024enhancingradioisotopeidentificationgamma} approaches to background estimation and anomaly detection have also been developed.

    Though, in general, more complex algorithms seem to obtain better performance, recent research in machine learning and artificial intelligence has shown that even for sophisticated models there is an inevitable drift over time away from optimal performance due to the widening gap between the data used to train the model and the current data on which it is being evaluated~\cite{drift1, drift2, drift3}. Model drift is closely related to point~\ref{rapid} above, and effective anomaly detection algorithms should have methods in place to combat it.
    
    Non-negative matrix factorization (NMF)~\cite{lee_learning_1999} has been shown to be a useful framework for modeling gamma-ray background signatures~\cite{aerial1, panda1, bilton_nonnegative_2019, bandstra_correlations_2021} when using ``global'' models trained on large datasets containing data representative of background signatures present in test cases. However, the existing Berkeley Lab-developed NMF-based approach~\cite{bilton_nonnegative_2019} lacks the flexibility required for deployment on mobile detector systems in scenarios where a pre-trained model is unable to capture the complexity of unseen data. In Section~\ref{sec:nmf1}, we will discuss why NMF-based algorithms are inherently well suited for detection and identification tasks and how the constraints of the conventional approach lead to clear limitations of performance on mobile detection tasks. We then present Adaptive NMF in Section~\ref{sec:adaptive}, an update to the conventional approach that seeks to address the issue of temporal drift and extend the NMF framework to unseen environments. 
    
    Finally, Section~\ref{sec:results} we test the detection performance of Adaptive NMF against a suite of baseline algorithms including gross counts k-sigma~\cite{jarman_comparison_2008, archer_systematic_2015}, multiplexed censored energy window (mCEW)~\cite{mcew}, the "conventional" NMF approach~\cite{bilton_nonnegative_2019}, and Nuisance-Rejection Spectral Comparison Ratio (NSCRAD)~\cite{detwiler_spectral_2015, pfund_improvements_2016}. \footnote{These algorithms were selected as benchmark comparisons because they are commonly fielded in real-world applications, there exist open-source implementations, and were recommended by the Detecting Radiation Algorithms Group~\cite{drag}. This is by no means an exhaustive list of algorithms used for anomaly detection tasks, and there exist methods used by entities in private industry. However, due to the proprietary nature of such methods, we were unable to test against them.} We show that Adaptive NMF achieves anomaly detection performance superior to  the benchmarks while maintaining a low FAR under rapidly changing environmental conditions. 
    
    The algorithm was developed and tested using a combination of synthetic and experimental data from mobile detection systems. The benchmark algorithms against which Adaptive NMF is tested are implementations based on their descriptions in the literature from the Radiological Anomaly Detection and Identification (RADAI) project~\cite{radaiRepo}.

\section{Non-negative Matrix Factorization}
    \label{sec:nmf1}
    The notation used in the following presentation of NMF closely follows~\cite{bilton_nonnegative_2019} and~\cite{panda1}.
    \subsection{Background Modeling with NMF}
        NMF is a dimensionality reduction technique which takes a data matrix $\textbf{X} \in \mathbb{R}_{\geq0}^{n \times d}$ and decomposes it into a product of two low-rank matrices 
        \begin{equation}
        \textbf{X} \approx \hat{\textbf{X}} = \textbf{A}\textbf{V},
          \ \ 
        \textbf{A} \in \mathbb{R}_{\geq0}^{n \times k}, \ \ 
        \textbf{V} \in \mathbb{R}_{\geq0}^{k \times d},
        \end{equation}
        where $d$ is the dimension of each of the vectors, $n$ is the number of vectors in the dataset, and $k$ is the rank of the resulting approximation matrix, $\hat{\textbf{X}}$. The columns of the matrix $\textbf{V}$ are a set of basis vectors for a lower-dimensional space that contains the Poisson expectation values of the original training data points\footnote{The space is not a true vector space because of the non-negative constraints. In particular, the set more closely resembles a conical hull with the additional constraints that all vectors in the set be non-negative.}. In practice, the basis vectors can be interpreted as background spectral templates developed from the training process. The rows of $\textbf{A}$ function as the coefficients for linear combinations of the basis vectors that approximate the members of the original dataset. By decomposing a set of gamma-ray spectra in this way, the columns of $\textbf{V}$ can be interpreted as the constituent components of the local gamma-ray background signature. In order to force the NMF components to be consistent with Poisson statistics, the background model is fit by optimizing the Poisson negative log likelihood loss function:
        \begin{equation}
        \label{eq:loss}
            {\cal L} = - \log L(\mathbf{X} | \mathbf{A}, \mathbf{V})
                    \equiv \sum\left(\mathbf{A}\mathbf{V}
                                    - \mathbf{X} \odot \log(\mathbf{A}\mathbf{V})
                                \right)
        \end{equation}
        where \( \odot \) is the element-wise matrix product and the sum is taken over the entire matrix. In equation~\ref{eq:loss}, the matrix $\textbf{X}$ is now also constrained to have nonnegative integer-valued entries, representing the counts in each gamma-ray spectral bin.
        
        The non-negativity constraint of NMF admits background models that are more consistent with photon physics than other dimensionality reduction methods such as singular value decomposition (SVD). Consequently, NMF models lend themselves well to physical interpretability and potential correlation with real features of the environment without the addition of explicit information about nuclear physics~\cite{panda1, aerial1}. While increased interpretability is often observed in practice, NMF does not guarantee interpretability for the following reasons:
        \begin{itemize}
            \item \textbf{NMF has no closed form solution}: NMF is computed by minimizing a loss function via maximum likelihood expectation maximization.
            \item \textbf{NMF solutions are not unique}: NMF requires an initial guess, which is often randomly initialized, so there is no guarantee that two solutions trained on the same dataset will admit the same decomposition if they are instantiated differently. Furthermore, one could construct a positive-definite linear transformation $\textbf{W}$, apply it to $\textbf{A}$ and its inverse to $\textbf{V}$, and be left with a distinct and equally valid decomposition, as: $
            \textbf{X} \approx \hat{\textbf{X}} = \textbf{A}\textbf{V} = \textbf{A}(\textbf{W}\textbf{W}^{-1}) \textbf{V} = (\textbf{A}\textbf{W})(\textbf{W}^{-1} \textbf{V}) \equiv \textbf{A}^\prime\textbf{V}^\prime $
            
        \end{itemize}
    
    Another important difference between NMF and methods like SVD is that the rank, $k$, of the approximation must be set prior to calculation. There exist many methods by which to choose an appropriate number of NMF components for a given training dataset. In this work, multiple models are trained with different values of $k$ and the Akaike Information Criterion~\cite{akaike} is used as a heuristic for choosing the best model.

    \subsection{Anomaly Detection and Isotope Identification}
        \label{sec:detection}
        In prior work, NMF has been used to perform anomaly detection and isotope identification using the following protocol:
        
        \begin{enumerate}
            \item Reconstruct an incident spectrum with the NMF background model and call its loss value $\mathcal{L}_B$.
            \item For each source template in a template library $t_i$, add the template to the background model and reconstruct again. Call its negative Poisson log likelihood loss $\mathcal{L}_{t_i}$.
            \item Perform a likelihood ratio test (LRT) for each of the sources in the library. This test statistic will function as the alarm metric $\ell_{t_i} \equiv 2 (\mathcal{L}_B- \mathcal{L}_{t_i})$.
            \item If any of the $\ell_{t_i}$ is above a precomputed threshold, raise an alarm. Take the argument maximum over all of the $\ell_{t_i}$ to be the isotope identification inference.
        \end{enumerate}
        The distribution of the LRT statistic should roughly follow a $\chi^2$ distribution with a single degree of freedom, but because the statistic is clipped at zero, the distribution will on average have about half of its values at zero. Thresholds can be set to yield a desired FAR by calculating the value where the integral of the $\chi^2$ tail distribution is twice the desired probability of false alarm (PFA).

        \begin{figure}[h!]
            \hspace{-0.3in}
            \includegraphics[width=0.52\textwidth]{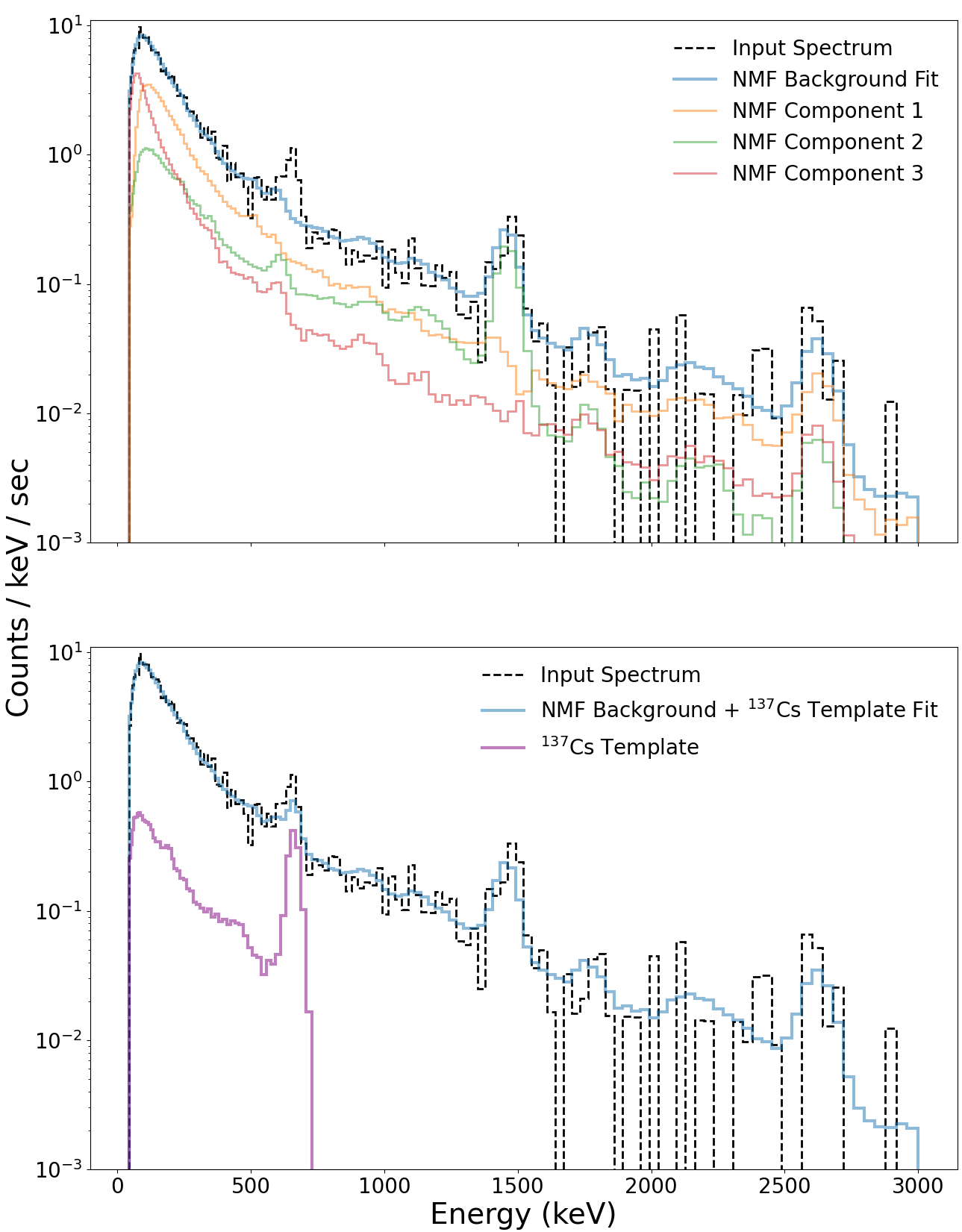}
            \caption{A \ce{^137Cs}-containing spectrum with the NMF background model fit and individual components (top), and the background+source template fit with the \ce{^137Cs} template (bottom). NMF components and \ce{^137Cs} template have been scaled by their corresponding reconstruction weights. \label{fig:fit}}
        \end{figure}

        \begin{algorithm}[h!]
            \caption{Conventional NMF Analysis}\label{alg:nmf}
            \begin{algorithmic}[1]
                \State \textbf{Require}: $\textbf{x}$, incident spectrum
                \State \textbf{Require}: $\mathcal{T}$, detection threshold
                \State $\hat{\textbf{x}}_B \gets \mathrm{fit}_{B}(\textbf{x}) $
                
                \State $\mathcal{L}_B \gets \mathrm{NPLL}(\textbf{x}, \hat{\textbf{x}}_B)$ \Comment{Neg. Poisson log likelihood}
                \For {$t_i \in \{t_0, t_1, ..., t_m\}$} \Comment{Iterate through templates}
                    \State $\hat{\textbf{x}}_{t_i} \gets \mathrm{fit}_{B+t_i}(\textbf{x})$
                    \State $\mathcal{ L }_{t_i} \gets \mathrm{NPLL}(\textbf{x}, \hat{\textbf{x}}_{t_i})$
                    \State $\ell_{t_i} \gets \max(2(\mathcal{L}_B-\mathcal{L}_{t_i}), 0)$ \Comment{Likelihood Ratio Test}
                \EndFor
                \State $is\_alarm \gets \max(\ell_{t_i} | \forall i \in \{0, ..., m\}) > \mathcal{T}$
                \State $id\_inference \gets \argmax(\ell_{t_i} | \forall i \in \{0, ..., m\})$
            \end{algorithmic}
        \end{algorithm}
        
        The algorithm set forth above (also shown in pseudocode in Algorithm~\ref{alg:nmf}) will be referred to henceforth as `Conventional NMF.' This method of analysis runs at a rate of approximately 60 iterations per second on a 10-core Apple M1 chip. Similar performance was observed on an Intel\textcircled{\footnotesize R} Core\textsuperscript{TM} i7-1370P 14-core Processor. An example of a reconstruction of a spectrum containing \ce{^137Cs} using NMF background components and a \ce{^137Cs} template is shown in Fig.~\ref{fig:fit}, adapted from \cite{bilton_nonnegative_2019}. In this example, the addition of the source template improves the quality of the reconstruction and yields an alarm metric value $\ell_{^{137}{Cs}} = 36.8$ significantly above the detection threshold $\mathcal{T} = 21.6$.

        Conventional NMF has been shown to perform well on detection and identification tasks when trained on a large amount of data representative of the testing environment. However, if this requirement is not met, as is often the case on mobile detector systems, the background model will no longer effectively represent the current environment as the detector enters unfamiliar territory, and detection sensitivity will suffer. Here we introduce `Adaptive NMF', a more generalized approach to NMF-based anomaly detection and isotope identification whose purpose is to overcome the limitations of Conventional NMF.

\section{Adaptive NMF}
    \label{sec:adaptive}
    The object of the Adaptive NMF algorithm is to maintain or exceed the detection sensitivity offered by the conventional NMF-based approach, but with the ability to adapt to changing environments and deployment modes without the need for a system operator to manually tune hyperparameter values, retrain the background model, or restart the algorithm altogether. In order to do so, the algorithm continuously collects background gamma-ray spectra and uses them to periodically refit its background model to remain consistent with local background conditions. The core functionality (and by extension the computational performance) of the algorithm is identical to Algorithm~\ref{alg:nmf} with the following modification to the logic in the analysis step:
    \begin{itemize}
        \item Analyze the current spectrum with the most up-to-date background model
        \item If the algorithm does not flag the spectrum as anomalous, add it to a rolling buffer of background data to be used for retraining
        \item Refit the background model periodically.
    \end{itemize}
    The following sections describe additional features implemented in the Adaptive NMF framework to improve its detection accuracy and sensitivity.

    \begin{figure*}[htb!]
        \begin{centering}
        \includegraphics[width=\textwidth]{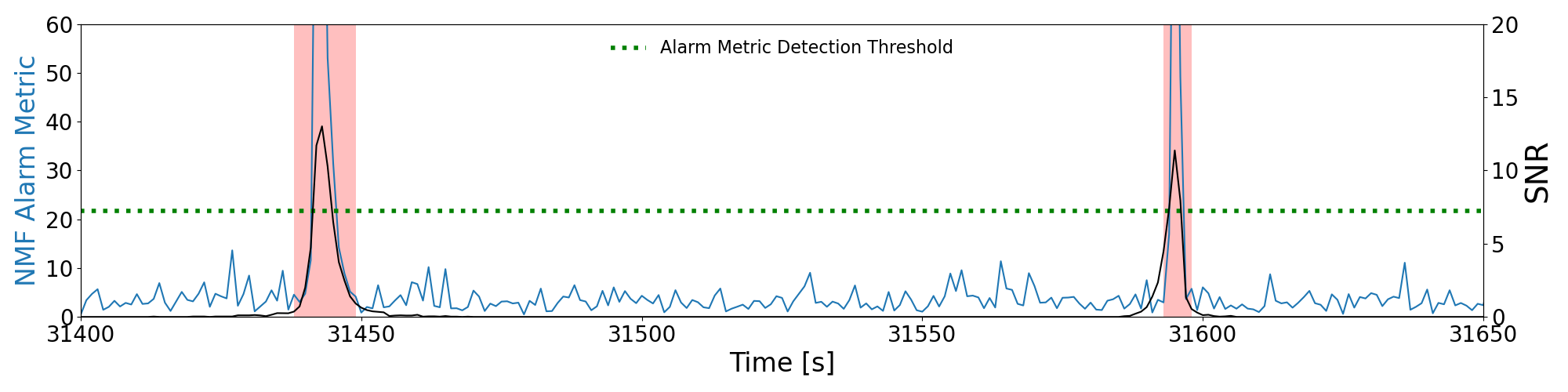}
        \caption{Adaptive NMF alarm metric values and inferred encounter windows (shown in red) using the temporal windowing method. 
         \label{fig:TW_AM}}
        \end{centering}
    \end{figure*}
    
    \subsection{Automatic Refitting}
    The adaptive NMF algorithm must be able to retrain its background model while maintaining real-time performance (typically operating on incident spectra at 1--2\,Hz). The model retraining step was implemented as a parallel process which waits idly until it receives a signal to begin. The background buffer on which the NMF model is trained contains the most recent hour of spectra (3600~spectra at 1~hz) that did not raise alarms, and retraining takes place every 10~minutes. Both the buffer length and the retraining frequency are adjustable parameters that can be tuned based on a user's specific practical or computational needs. 
    
    When collecting spectra from large-volume (e.g., 2"$\times$4"$\times$16" NaI) detectors in mobile operations, we found that three NMF components are typically sufficient to describe the background. Such models take about 20~seconds on average to retrain. In practice, this translates to approximately 3\% of spectra being analyzed with the previously trained background model. No case in which this behavior leads to a degradation of detection accuracy or sensitivity has been observed in practice.

    \subsection{Tikhonov Regularization}
    \label{sec:tik}
    Gamma ray signatures of NORM sources are spectroscopically similar to the background which can lead to a mathematical degeneracy during analysis. In order to suppress the effects of this degeneracy --- namely, false alarms arising from NORM anomalies --- a Tikhonov regularization penalty was applied to the alarm metric value during analysis. This regularizer is functionally equivalent to applying an L$_2$ penalty only to the source template component's coefficient. This forces the algorithm to preferentially use the background model components over the source templates during reconstruction. As a result, the more distinct a source template's spectral shape is from the background components, the smaller the regularizer's effect on the template's coefficient will be, and vice versa.
    
    The addition of this regularizer greatly improves the consistency of the desired and observed false-alarm rates. However, one drawback of this method is that the algorithm loses sensitivity to NORM in the environment. In practical applications, NORM is often not regarded as a threat source. Therefore, in the context of an urban search task, the benefits of this method outweigh the risks associated with decreased sensitivity. The strength of the regularizer is also an adjustable parameter that can be tuned based on a user's specific needs.

    \subsection{Temporal Windowing}
    \label{sec:window}
    Because the Adaptive NMF algorithm is allowed to construct its own training datasets using uncontrolled, unlabeled data, there is a danger that signatures associated with weak sources (i.e., those that did not trigger an alarm) may be inadvertently incorporated into the background model. This would reduce the algorithm's sensitivity to similar source signatures in the future, as they would be more easily described by the background model. Therefore, mitigating false negatives is critical in maintaining detection performance. This is addressed by implementing a subalgorithm which constructs temporal exclusion windows around NMF alarms. In practice, we expect there to be periods during which a source is present but the signal strength is too weak to be detected. The temporal windowing method aims to prevent these low-SNR source-containing spectra found at the beginnings and ends of encounters from entering the background buffer. 

    
    When a spectrum raises an alarm, a window containing only the alarming spectrum is constructed. The algorithm then iterates forwards and backwards in time, integrating the window as it goes, and analyzes the resulting multi-second spectrum. This process continues until the alarm metric of the integrated window falls below a secondary threshold, set as a fraction of the primary detection threshold. This fraction is a tunable parameter, with higher values corresponding to shorter windows, and vice versa. 
    
    In addition to preventing algorithm sensitivity loss, the temporal windowing mechanism also provides users with an estimate of the bounds of source encounters. It handles asymmetric encounter windows and encounters of different lengths, as shown in Fig.~\ref{fig:TW_AM}. The blue curve shows the algorithm alarm metric value over two separate simulated source encounters, the black curve shows the actual value of SNR (calculated as $\frac{s}{\sqrt{s+b}}$ with $s$, the number of photons originating from the source and $b$, the number of photons coming from the background) at each second, and the red boxes are the exclusion windows determined by the temporal windowing logic. The subalgorithm successfully excludes multiple spectra with nonzero SNR from the background buffer that did not raise alarms.

    Another potential approach to mitigating cascading sensitivity loss due to the incorporation of source-containing spectra into the background buffer would be the addition of a secondary detection threshold, lower than the primary detection threshold. One could then exclude spectra from the background buffer whose maximum alarm metric exceeds this threshold even if the primary detection threshold is not exceeded. We did not pursue this approach due to its potential to filter out true background spectra with high variability. This would impede the background buffer from capturing some background fluctuations that are important for developing the NMF background model. The implementation of such an approach and comparing its efficacy to the temporal windowing method described here is a potential avenue for future work.
    
    \begin{figure}[h!]
        \hspace{-0.15in}\includegraphics[width=0.52\textwidth]{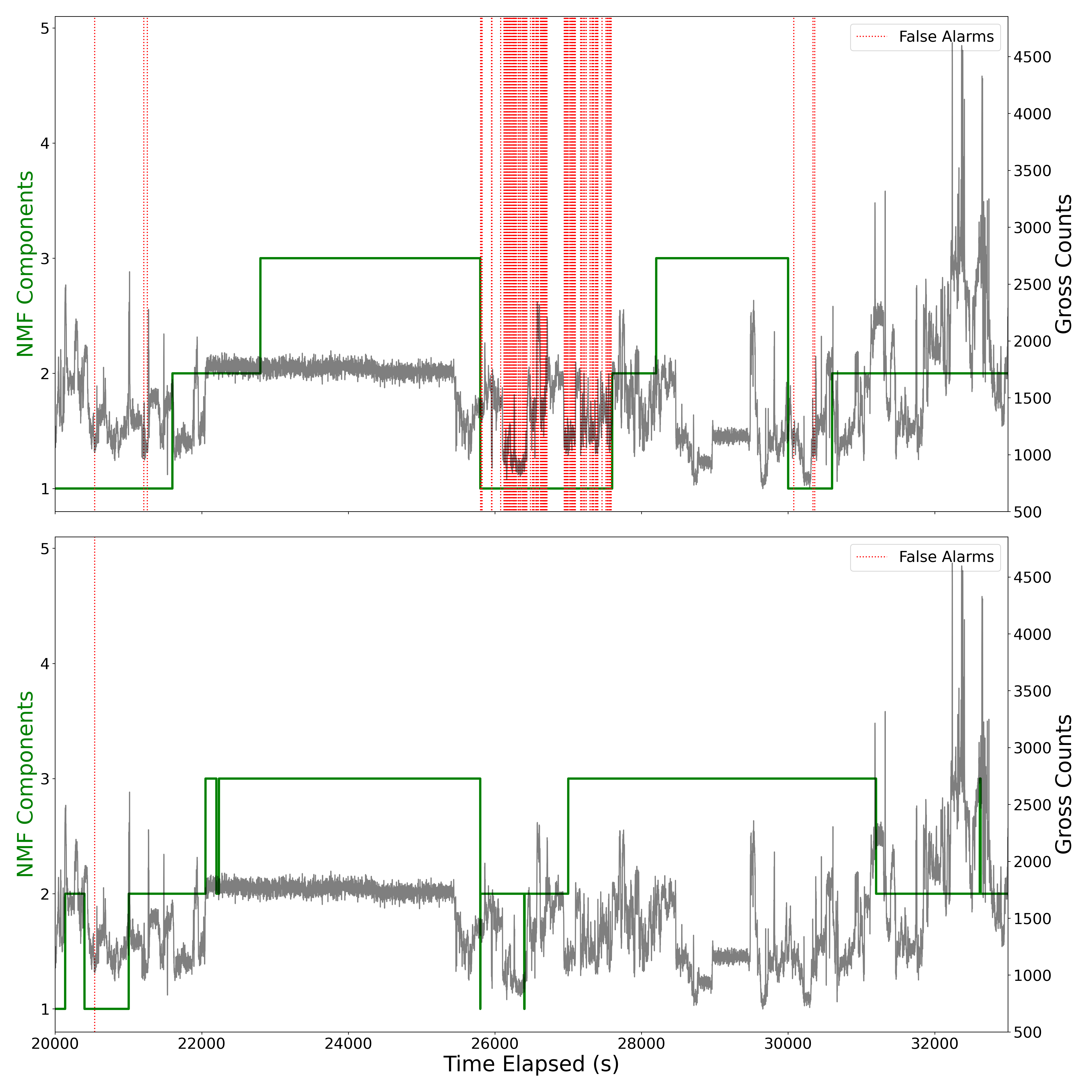}
        \caption{Number of NMF components being used at each second of analysis (green), gross count rate (black), and false alarms (red). Running Adaptive NMF with the model switching logic applied to analysis (bottom) yields far fewer false alarms than without (top). \label{fig:switch}}
    \end{figure}

    \begin{figure*}[htb!]
        \hspace{-0.25in}\includegraphics[width=1.1\textwidth]{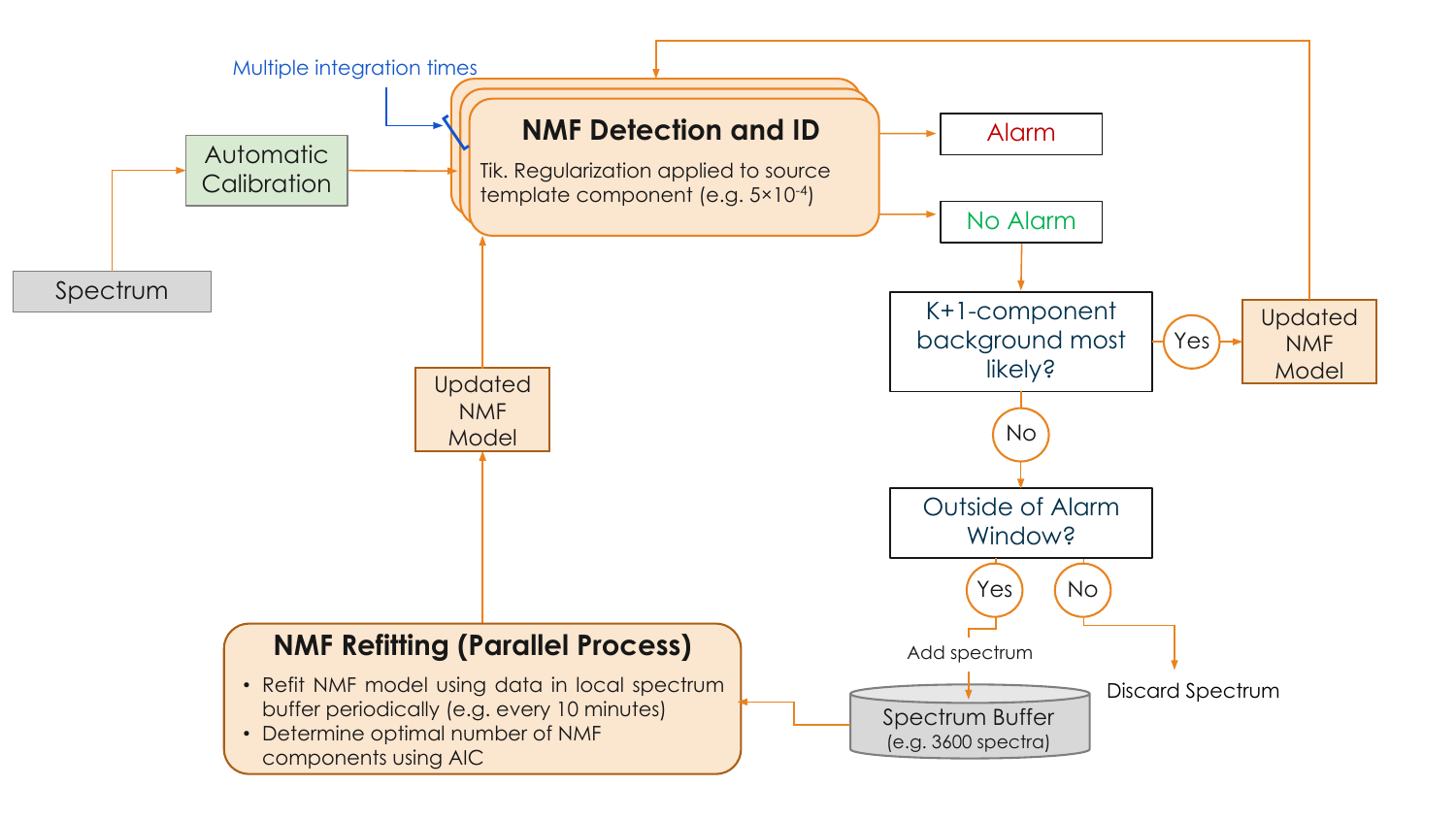}
        \caption{The complete system architecture diagram of the Adaptive NMF algorithm. \label{fig:sys}}
    \end{figure*}
    
    \subsection{Background Complexity Detection}
    \label{sec:switch}
    The ability of an algorithm to detect abrupt changes in the background and adapt accordingly is crucial for mobile search applications. Consider a scenario in which a detector system is at rest for an extended period of time (e.g., longer than the algorithm background buffer). Because gamma-ray backgrounds are less variable when observed on a static detector, the background buffer is populated with low-variability spectra, and Adaptive NMF will select a simple model. If the detector suddenly began to move, Adaptive NMF would likely raise many false alarms due to the complexity of the current (static) background model being insufficient to describe the dynamic and changing environment. 

    In order for the algorithm to detect an abrupt change and adjust its background model accordingly, an additional step was added to the analysis method. Namely, previous background models of varying complexity are kept in memory, and the algorithm performs an additional likelihood ratio test between the current $k$-component background model and the most recent $(k+1)$-component model. If this LRT statistic is the highest one (i.e., a more complex background is the most likely explanation for the incident spectrum), the background model is replaced with the more complex one. This process is repeated until the background-background LRT statistic, $\ell_{B_{k+1}} \equiv 2(\mathcal{L}_{B_k}-\mathcal{L}_{B_{k+1}})$, is no longer the highest value. 

    Fig.~\ref{fig:switch} shows the efficacy of this approach in suppressing false alarms that would otherwise have been raised by the algorithm. The detector is static for approximately one~hour, from approximately $t=22000 - 25600$, after which point the gross count rate, shown in black, becomes highly variable. The detector remains mobile for the remainder of the experiment, with the exception of a short period at $t \approx 29000$. The number of components Adaptive NMF uses for its background model at each second is shown in green. When the algorithm is allowed to switch its background to a more complex previously-used model upon encountering a sharp change in the background, it is able to avoid becoming stuck in a state of false alarm.

    \subsection{Multiple Integration Times}
    Because the optimal integration window to maximize the SNR of a given source encounter is unknown when running in arbitrary environments, multiple Adaptive NMF instances are run in parallel with varying integration times. We found that using three integration timescales (short: 1--3\,s, medium: 4--6\,s, and long: 7--10\,s) provided sufficient flexibility for the algorithm to detect source encounters of different lengths. Alarm metric values from each subprocess are concatenated, and the argument maximum overall is reported as the isotope identification inference label. 
    
    It was empirically observed that models derived from 1-second integrated spectra could be used to perform inference on multi-second integrated spectra with no noticeable loss in performance. For this reason, all background models used by the individual subprocesses are trained on 1-second integrated spectra. Each subprocess also performs model switching independently of the others, providing additional flexibility. Analysis time increases approximately linearly with the number of subprocesses employed. 
    
    \subsection{Automatic Energy Calibration}
    In order to account for gain shifts and other detector effects, an automatic energy calibration routine~\cite{marco1, marco2} was applied to each incoming spectrum before being sent to Adaptive NMF for analysis. The algorithm fits background spectra to full-spectral templates to perform continuous calibration. The method was tested on a NaI detector of a similar volume to those used throughout this work that experienced temperatures between -40ºC and +50ºC, during which time less than 0.4\% peak variability was observed at 122\,keV and less than 0.2\% at 1460\,keV. Fig.~\ref{fig:sys} shows the logical flow of the full Adaptive NMF algorithm with all of the features described in the preceding subsections.
    
\section{Results}
\label{sec:results}
    Adaptive NMF was tested against gross counts k-sigma~\cite{jarman_comparison_2008, archer_systematic_2015}, mCEW~\cite{mcew}, Conventional NMF~\cite{bilton_nonnegative_2019}, and NSCRAD~\cite{detwiler_spectral_2015, pfund_improvements_2016}. The implementations of the benchmark algorithms come from the RADAI algorithm library~\cite{radaiRepo}. 

    Analysis was performed by running sequential one-second integrated gamma-ray spectra through each algorithm. If there was any alarm within the ground-truth bounds of each encounter, it was counted as a detected encounter. Any alarm outside the bounds of any encounter was counted as a false alarm. Because only a subset of the benchmark methods are capable of performing isotope identification, only anomaly detection performance will be reported in this section.

    \begin{table}[t!]
        \caption{  \label{radai_sources}}
        \centering
        \begin{tabular}{ |p{1.75cm}|p{1.75cm}|p{1.75cm}|p{1.75cm}|  }
        \hline
         \multicolumn{4}{|c|}{Simulated Sources in RADAI Dataset}   \\
         \hline
         NORM& Industrial &Medical &(S)NM                           \\
         \hline
         \ce{^40K}   & \ce{^192Ir} & \ce{^131I}    & NatU           \\
         \ce{^232Th} & \ce{^60Co}  & \ce{^18F}     & HEU            \\

         \ce{^226Ra} & \ce{^133Ba} & \ce{^{99m}Tc} & LEU            \\
                     & \ce{^137Cs} & \ce{^57Co}    & Refined U      \\
                     &             & \ce{^201Tl}   & \ce{^241Am}    \\
                     &             & \ce{^90Sr}    & FGPu           \\
                     &             & \ce{^67Cu}    & WGPu           \\
         \hline
        \end{tabular}
        \vspace{5pt}%
        
        (HEU:~High-Enriched Uranium, LEU:~Low-Enriched Uranium, NatU:~Natural Uranium, FGPu:~Fuel-Grade Plutonium, WGPu:~Weapons-Grade Plutonium)
    \end{table}

    \subsection{Simulated Datasets}
    RADAI provides a dataset~\cite{radai_dataset_1, radai_datasets_webpage} that simulates a 2"$\times$4"$\times$16" NaI detector moving through an urban environment. The purpose of the RADAI dataset is to provide users with simulated source encounters in environments where data collection would be prohibitively expensive or otherwise infeasible. It includes sources one might typically encounter in an urban environment like those used in medical and industrial applications, threat sources like highly enriched Uranium and weapons-grade Plutonium, and NORM. All sources also have varying shielding configurations common for their individual applications. Single-source encounters with all 21 of the sources listed in Table~\ref{radai_sources} are included in the dataset. The data were prepared by integrating list-mode events into a series of one-second spectra in 129 nonlinear energy bins from 45--3000\,keV. The width of these bins is proportional to the square root of the energy in order to ensure that the ratio of the bin widths to the detector energy resolution is approximately constant throughout the entire spectrum. 
    The template library used by each of the benchmark algorithms that make use of spectroscopic information (i.e., all but k-sigma) was also generated from RADAI source data.
    
    For these tests, Adaptive NMF was instantiated with a background buffer length of 10000~1-second integrated spectra; a refit period of 600~seconds (10~minutes);  integration times of 1, 2, 3, 4, and 5~seconds running in parallel; and a Tikhonov regularization factor of $10^{-4}$. The Conventional NMF instance was trained on 10~hours' worth of background data (36000 spectra).

\begin{figure}[t!]
        \hspace{-0.15in}\includegraphics[width=0.52\textwidth]{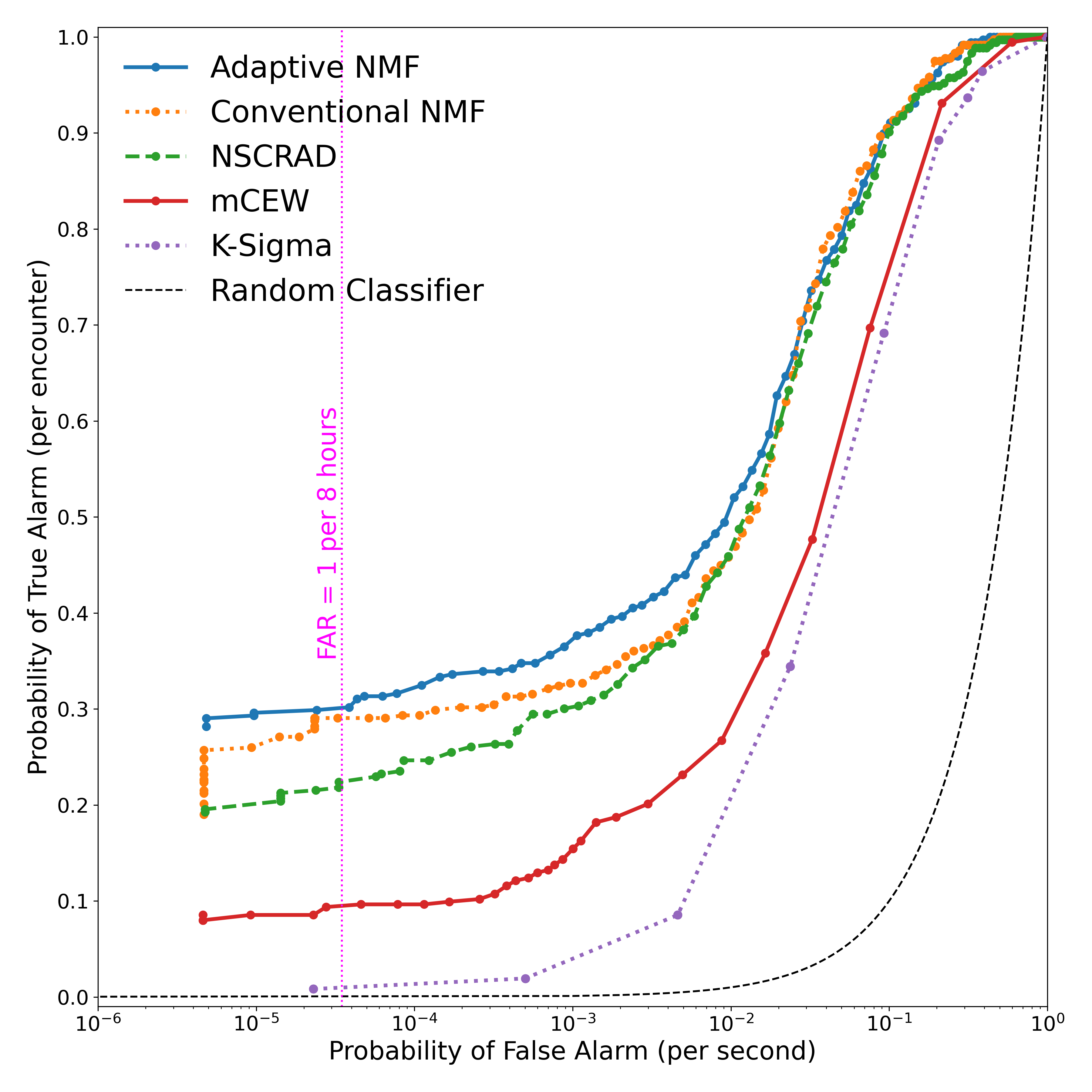}
        \caption{A set of ROC curves comparing the detection performance as a function of false alarm rate for each of the benchmark algorithms and Adaptive NMF. Adaptive NMF performs the best at functional false alarm rates. Note the log scale on the x-axis.\label{fig:rocFull}}
    \end{figure}

    
    Fig.~\ref{fig:rocFull} shows a set of Receiver Operating Characteristic (ROC) curves comparing the detection performance of Adaptive NMF against the suite of RADAI benchmark algorithms tested on all 21~simulated sources. The x-axis of the ROC curve is the PFA per second, and the y-axis gives the probability of detection (PD) per source encounter. Points on the curve are calculated by varying each algorithm's detection threshold value and observing the corresponding PD and PFA. All other algorithm hyperparameter values are fixed. At functionally relevant false alarm rates (e.g., 1~per 8~hours, corresponding to a false alarm probability of about $3.5 \times 10^{-5}$), Adaptive NMF outperforms all benchmark methods.
    
\begin{figure*}[htb!]
        \hspace{-0.1in}\includegraphics[width=1.02\textwidth]{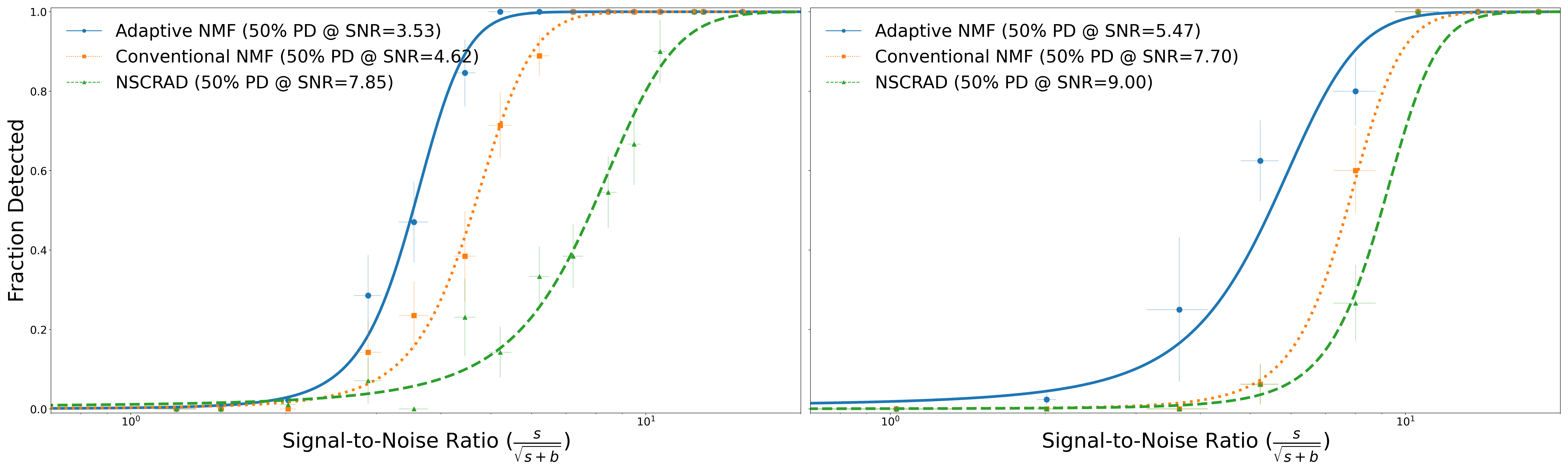}
        \caption{Logistic regression-fit sigmoids representing the probability of NSCRAD (green), Conventional NMF (orange), and Adaptive NMF (blue) as a function of SNR at a fixed FAR (1 per hour) for \ce{^60Co} (left) and \ce{^241Am} (right) encounters. The further a curve is to the left, the better the algorithm is at detecting weak sources. Adaptive NMF achieves a 50\% probability of detection on both sources at a lower SNR than either of the other two algorithms. Error bars show one-sigma uncertainty in both the $x$~and~$y$ axes. \label{fig:sigmoids}}
    \end{figure*}

    Fig.~\ref{fig:sigmoids} shows the performance (i.e., PD as a function of SNR at a fixed PFA) of Adaptive NMF against Conventional NMF and NSCRAD on two single-isotope detection tasks. Similarly, Adaptive NMF demonstrates superior detection performance over the other algorithms, achieving 50\% probability of detection at an SNR of 3.53 on \ce{^60Co} compared to 4.62 for Conventional NMF and 7.85 for NSCRAD. On \ce{^241Am} encounters, Adaptive NMF achieves a probability of detection of 50\% at an SNR of 5.47 compared to 7.70 for Conventional NMF and 9.00 for NSCRAD.

    \subsection{Field Tests}
    The following results come from tests performed on data collected in urban and rural locations with vehicle-borne 2"$\times$4"$\times$16" NaI detectors. We performed these tests with an instance of Adaptive NMF equipped with a background buffer of 3600~1-second integrated spectra; a refit period of 600~seconds (10~minutes); integration times of 1, 2, 4, and 8~seconds running in parallel; and a Tikhonov regularization factor of $5\times10^{-4}$. Adaptive NMF was tested against Conventional NMF (trained on a dataset of 8742 spectra representative of the testing environment) and an implementation of NSCRAD that has been used in field applications. These tests consisted of encounters with four different sources at six different distances. Eight passes were performed at each distance. 

\begin{figure*}[htb!]
        \includegraphics[width=\textwidth]{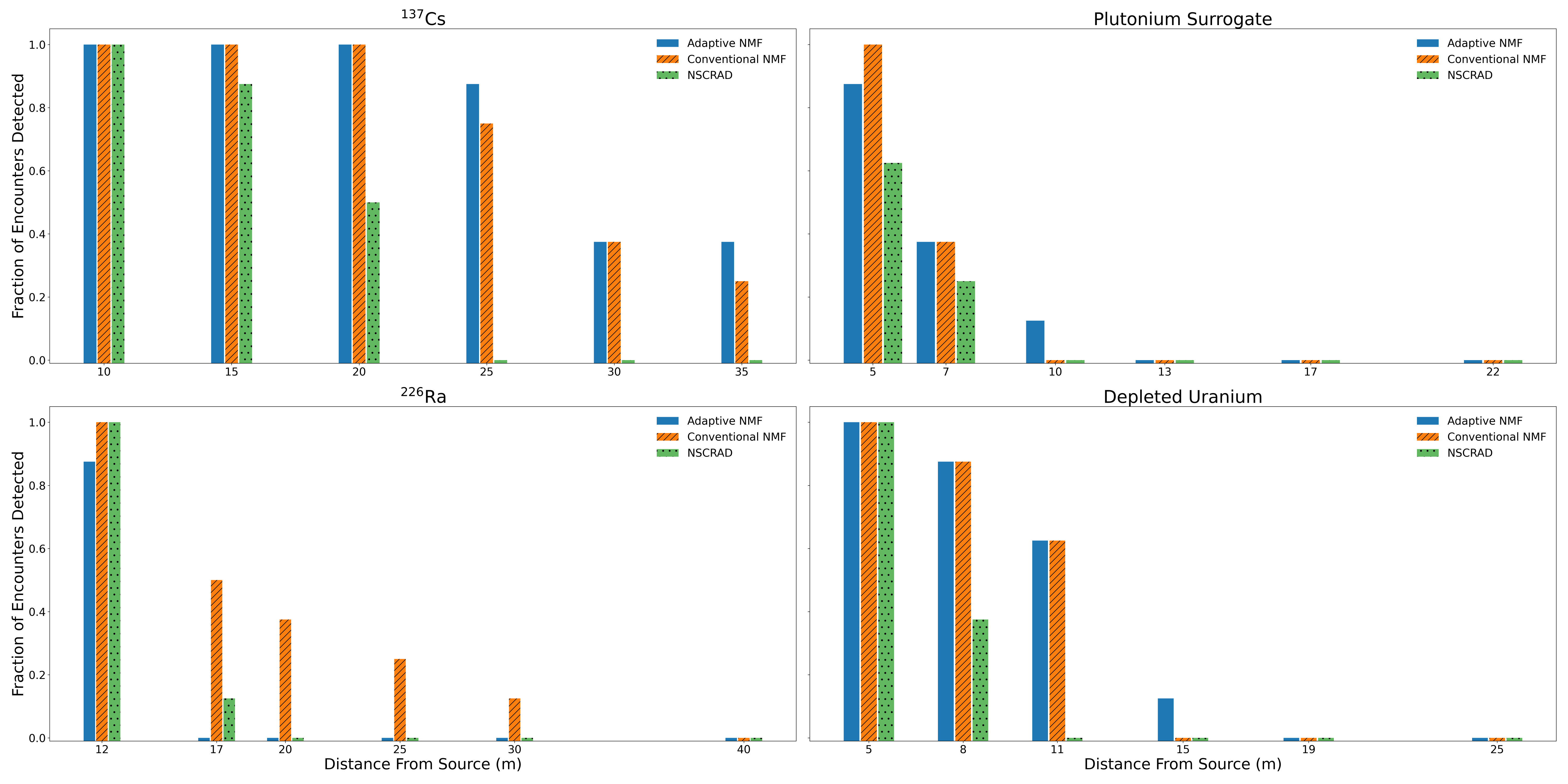}
        \caption{The fraction of real-world source encounters that were detected by Adaptive NMF (blue), Conventional NMF (orange), and NSCRAD (green) at each standoff distance at a fixed FAR of 1~per hour. \label{fig:vs_nscrad}}
    \end{figure*}

    Test results are shown in Fig.~\ref{fig:vs_nscrad}. Adaptive NMF is consistently able to detect non-NORM sources at larger standoff distances than NSCRAD and Conventional NMF at a fixed FAR of 1~per~hour. In the case of NORM sources like $^{226}$Ra, the detection performance of Adaptive NMF is poorer than that of the benchmark algorithms because such sources closely resemble background, and the Tikhonov penalty imposed during analysis (described in Section~\ref{sec:tik}) suppresses their alarm rates. As such, encounters with NORM sources are largely rolled into the background buffer and used for subsequent model refits, further exacerbating the difficulty in detecting them. However, while natural uranium, for example, is difficult to detect due to its high similarity with background spectral features, the ability of Adaptive NMF to detect uranium increases as the isotopic composition of the uranium is moved further away from secular equilibrium. Fig.~\ref{fig:vs_nscrad} shows that Adaptive NMF remains the most sensitive of the three algorithms on encounters with depleted uranium at large standoff distances.

    \begin{figure*}[t!]
        \includegraphics[width=\textwidth]{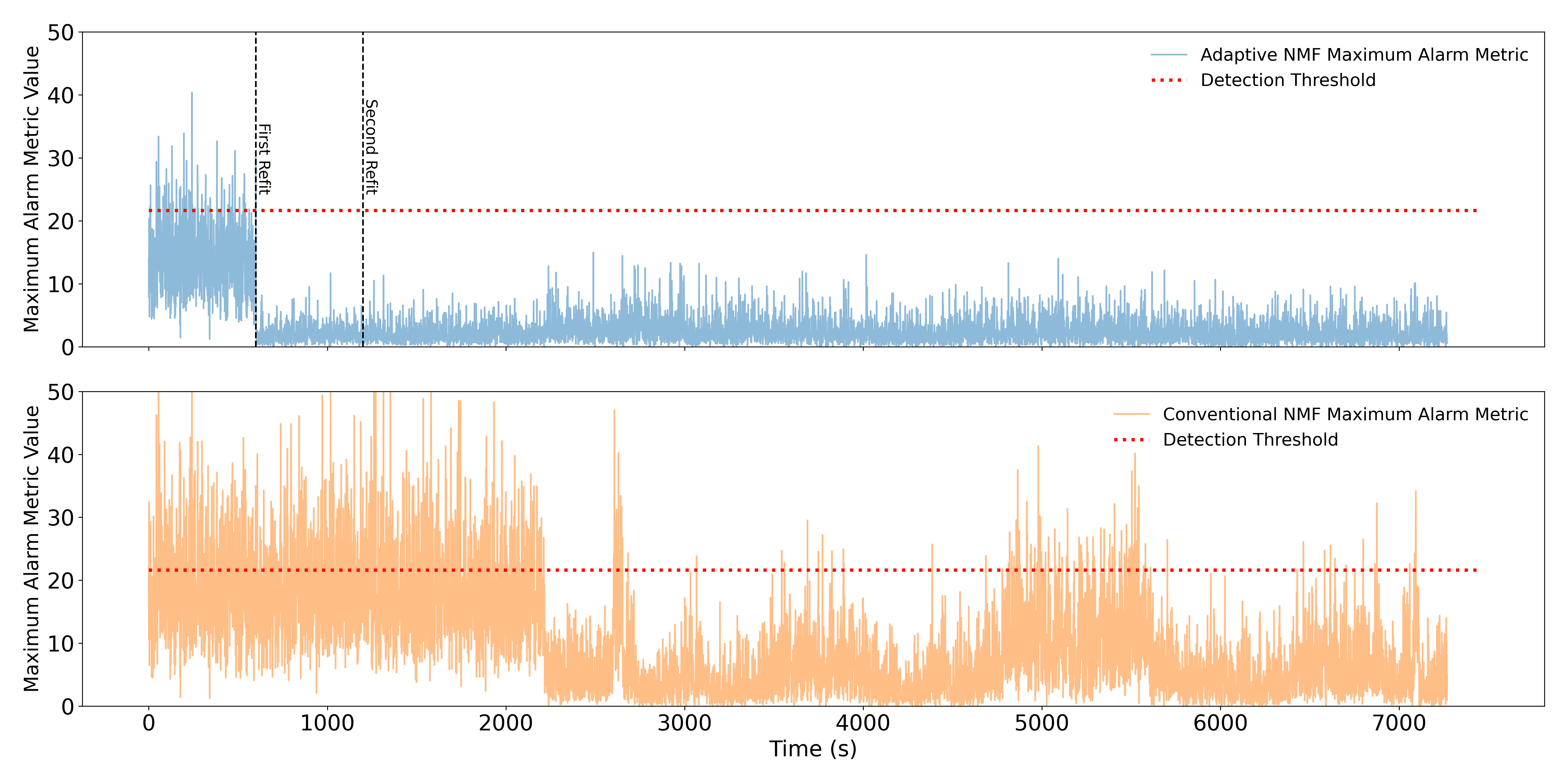}
        \caption{Conventional (bottom) and Adaptive (top) NMF alarm metric values when initially trained on a dataset of gamma-ray spectra collected in Washington D.C. and subsequently tested on spectra collected in rural New Mexico.  \label{fig:nmdc}}
    \end{figure*}

     \begin{figure*}[t!]
        \includegraphics[width=0.5\textwidth]{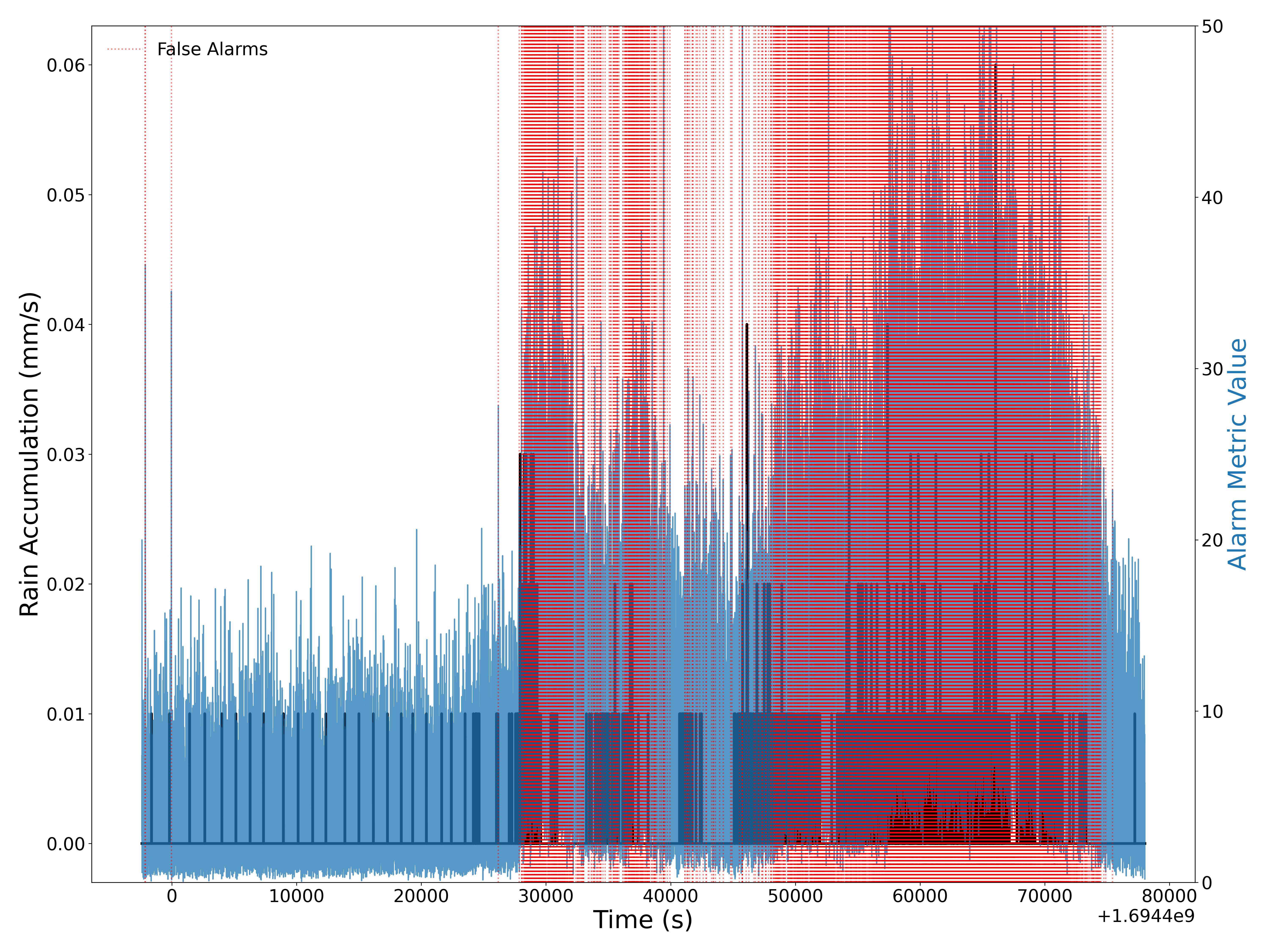}
        \includegraphics[width=0.5\textwidth]{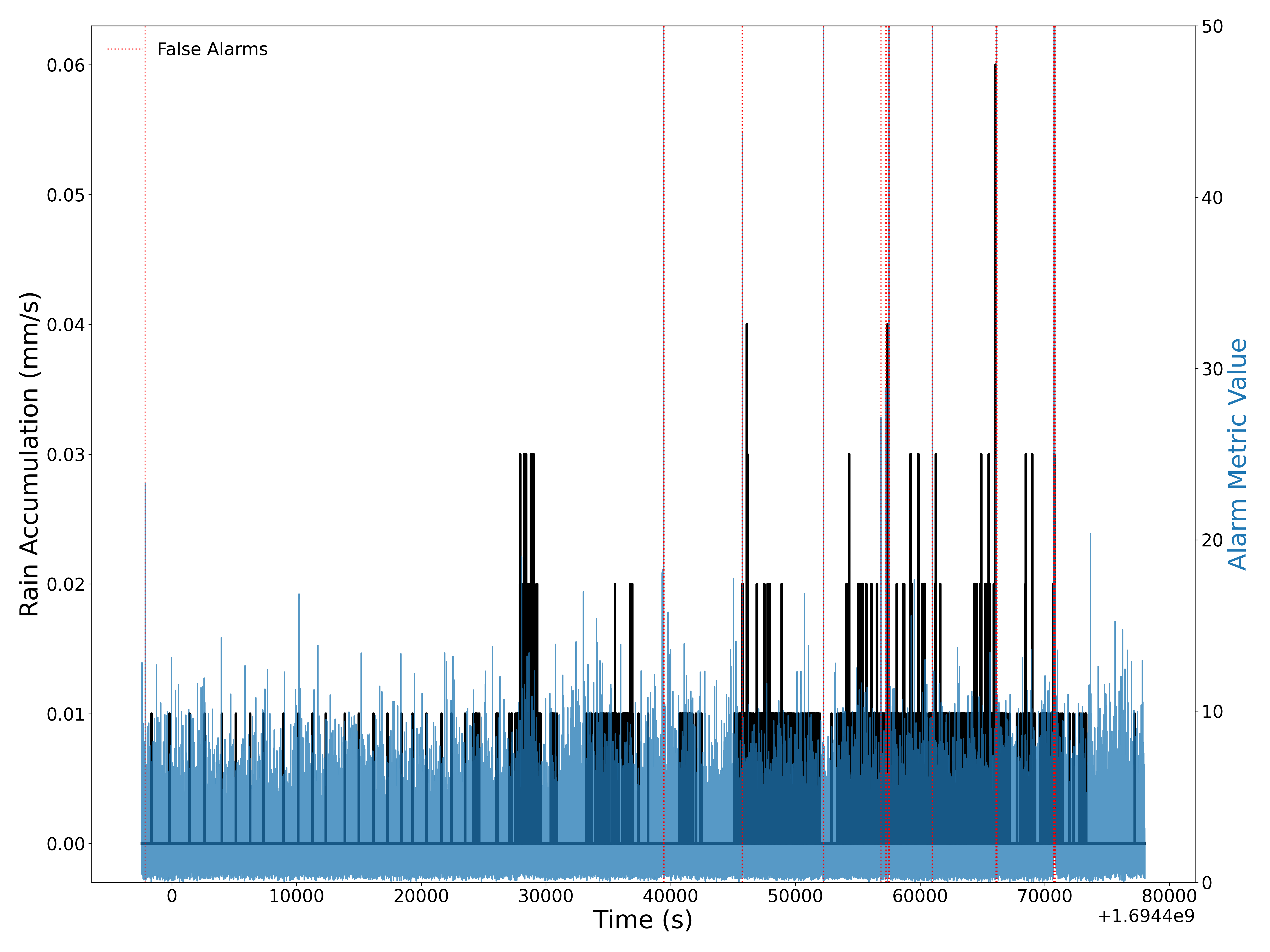}
        \caption{Conventional (left) and Adaptive (right) NMF alarm metric values over a rain event from the PANDAWN project. The blue curves are the maximum alarm metric values at each second, the black lines show the instantaneous accumulation value from rain sensor on the PANDAWN node, and the red dotted lines denote times at which the alarm metric exceeded the detection threshold, $\mathcal{T} = 21.6$, calculated as in Section~\ref{sec:detection}.  \label{fig:rainFA}}
    \end{figure*}
    
    \subsection{Nuisance Alarm Suppression}
    The Adaptive NMF algorithm is able to effectively adapt to changes in background environment and maintain a low and consistent false alarm rate compared to the conventional approach, as illustrated in Fig.~\ref{fig:nmdc}. Both methods were trained on a dataset of spectra collected in Washington D.C. and then tested on a dataset of spectra collected in rural New Mexico. Conventional NMF (bottom) raises numerous false alarms due to the inconsistency of its background model with the environment on which it is being tested. In contrast, Adaptive~NMF (top) begins analysis with a high false alarm rate as it gathers new background data, but is able to effectively capture the spectral signature in its background model after the first model refit. 
    
    Adaptive NMF has a clear advantage over the other benchmark methods --- particularly Conventional NMF --- during rain events. As it begins to rain, there is a gradual increase in radon progeny in the background~\cite{livesay_rain-induced_2014}. If a Conventional NMF instance faced such a scenario without having been previously trained on spectra collected during rain, it would likely raise false alarms as the concentration of radon exceeds the expected level. Without operator intervention, the false alarms would continue for hours after the rain stops until the radionuclides from the rain decay to a point where they no longer contribute to a significant component of the background. 

   Fig.~\ref{fig:rainFA} shows the Conventional and Adaptive NMF alarm metric values over a 25-hour period which included a rain event. These data were collected from a static detector node deployed as part of the PANDAWN project~\cite{Cooper_2023}. Adaptive NMF (right) was able to incorporate the background change into its model and raise far fewer false alarms than Conventional NMF (left) during the rain event.

\section{Discussion}
We have presented an NMF-based spectroscopic anomaly detection and isotope identification algorithm that generalizes more easily to unseen environments than existing comparable methods. In particular, Adaptive NMF met or exceeded the detection performance of Conventional NMF in most test cases. This is an encouraging result because Adaptive NMF is generally trained on far less data than Conventional NMF at any given point and makes fewer \textit{a priori} assumptions about the underlying environmental conditions. By demonstrating that a local model is sufficient for robust anomaly detection, we have extended the existing NMF framework to rapidly changing background conditions. 

We note the following potential weakness of the Adaptive NMF algorithm: As depicted in Fig.~\ref{fig:vs_nscrad}, our method has reduced sensitivity to NORM anomalies. The Tikhonov parameter was implemented to suppress the effects of the mathematical degeneracy arising from the spectroscopic similarity of NORM signatures with the background and to reduce false alarms arising from NORM anomalies. However, as a result, the values of the NORM template alarm metrics are suppressed, which makes it even more challenging to distinguish true signatures from typical Poisson fluctuations. The strength of the regularizer can be tuned or set to 0, giving users the flexibility to increase the algorithm's sensitivity to NORM, though it comes at the expense of the FAR. 

Perhaps unsurprisingly, the performance of the methods tested against the RADAI dataset, as shown in Fig.~\ref{fig:rocFull}, follows the same hierarchy as the amount of information being leveraged in each method. K-sigma uses only gross count rate information and had the poorest performance, which was to be expected in the presence of a rapidly changing background. mCEW, which constructs spectral regions of interest from its training data to perform detection, had greater sensitivity than k-sigma but still suffered when faced with the highly variable RADAI background. NSCRAD, too, constructs spectral windows to detect broad classes of anomalies, but it also incorporates information about its environment via a continually updating background model. This enables it to outperform mCEW and k-sigma, but it is susceptible to becoming stuck in a state of false alarm when abrupt background shifts occur. Conventional NMF achieved greater sensitivity by preserving the most spectral information compared to the other methods described and by making use of a broad library of templates of potential sources and a background model trained on many hours of data. The fact that Adaptive NMF was able to, in most test cases, match or improve upon the detection sensitivity of Conventional NMF is particularly encouraging due to the tradeoff in the information being utilized by each method. Adaptive NMF builds its background models from much smaller datasets --- on the order of one hour's worth of data --- while Conventional NMF commonly uses many hours of data to build a complex model that is capable of capturing the richness of an environment over time in a single model. We have shown that in the case of highly variable background compositions, a flexible local model is not only sufficient but oftentimes superior to a highly complex static model.

As we have demonstrated with our method, the NMF framework lends itself particularly well to the integration of additional features that further improve its detection performance. As such, there are many directions for future work that could further improve NMF-based methods. For instance, integrating an anomaly clustering method similar to that presented in~\cite{panda1} could allow Adaptive NMF to update its template library with observed anomalies, making it more sensitive to the specific shapes of source signatures in its environment. The inclusion of temporal information in the analysis pipeline could also increase detection sensitivity by allowing previous results to inform the algorithm's beliefs about future observations. Additionally, a continuously updating background model would generally be preferable to one that refits periodically to ensure that it does not become out-of-date between refits. A cursory exploration of such a method was performed, but it proved to be computationally intense and yielded no discernible difference in performance. However, it is possible that further investigation could uncover a feasible way to achieve a continuously updating model. 

\section*{Acknowledgments}
The authors wish to thank Jayson Vavrek and Brian Quiter for their helpful comments and suggestions for this manuscript.

\section*{CRediT statement}
\textbf{A.C.J.}:~Writing -- original draft, Formal analysis, Visualization, Software, Investigation (lead), Methodology.
\textbf{M.S.B.}:~Writing -- review \& editing, Supervision, Methodology, Formal analysis.
\textbf{S.F.}:~Investigation (supporting).
\textbf{Y.L.}:~Investigation (supporting).
\textbf{N.A.}:~Data curation.
\textbf{S.S.}:~Validation, Formal analysis, Data curation.
\textbf{R.J.C.}:~Writing -- review \& editing, Funding acquisition, Project administration, Supervision.

\bibliographystyle{IEEEtran}
\bibliography{adaptive_nmf}
\end{document}